\documentclass{INTERSPEECH2023}
\usepackage{bm,resizegather,cite}
\usepackage{footmisc}

\interspeechcameraready

\title{Target Speech Extraction with Conditional Diffusion Model}
\name{Naoyuki Kamo, Marc Delcroix, Tomohiro Nakatani}
\address{
  NTT Corporation, Japan}
\email{naoyuki.kamo.ka@hco.ntt.co.jp, marc.delcroix@ieee.org, tnak@ieee.org}

\begin{document}
\setlength{\abovedisplayskip}{2pt}
\setlength{\belowdisplayskip}{2pt}
\maketitle
 
\begin{abstract}
Diffusion model-based speech enhancement has received increased attention since it can generate very natural enhanced signals and generalizes well to unseen conditions. Diffusion models have been explored for several sub-tasks of speech enhancement, such as speech denoising, dereverberation, and source separation. In this paper, we investigate their use for target speech extraction (TSE), which consists of estimating the clean speech signal of a target speaker in a mixture of multi-talkers. TSE is realized by conditioning the extraction process on a clue identifying the target speaker. We show we can realize TSE using a conditional diffusion model conditioned on the clue. Besides, we introduce ensemble inference to reduce potential extraction errors caused by the diffusion process. In experiments on Libri2mix corpus, we show that the proposed diffusion model-based TSE combined with ensemble inference outperforms a comparable TSE system trained discriminatively.
\end{abstract}

\section{Introduction}
Speech enhancement consists of estimating clean speech signals from noisy recordings, which covers many sub-tasks such as noise reduction, dereverberation, speech separation, and target speech extraction (TSE). Speech enhancement research has made rapid progress with the advent of deep learning, leading to two main directions that differ by using discriminative or generative deep neural networks.

Discriminative approaches use a neural network trained to directly map noisy speech to clean speech by optimizing a signal level metric between the enhanced signal and a clean speech reference
 \cite{wang2018supervised}. 
These approaches are very powerful but sometimes lead to unpleasant artifacts and tend to be sensitive to mismatched conditions between training and testing. 

Generative approaches aim to model the distribution of clean speech using deep generative models \cite{pascual17_interspeech,Bando18_icassp,zhang21c_interspeech,lu22_iccassp,Welker22interspeech}, and exploit such speech priors to infer the clean speech from the noisy observation. Various deep generative models have been explored for speech enhancement, following their success in other fields. Among them, diffusion models have recently received increased attention since they have been shown to produce enhanced speech with high perceptual quality~\cite{Dolby-score-model}.  Moreover, they appear more robust to mismatch conditions between training and inference than discriminative models \cite{Julius22arxiv}. 

A diffusion model consists of a forward and reverse process \cite{YangSong-ICLR2021}. The \emph{forward process} transforms a data distribution into a known prior by adding noise to the data. The \emph{reverse process} reverses this transformation by gradually removing noise. It can be implemented using a neural network to predict scores (score model) and an iterative inference procedure called Langevin dynamics to generate samples. 
Diffusion model-based speech enhancement has been used to reduce various types of acoustic distortions such as background noise, reverberation, clipping, or codec artifacts~\cite{Dolby-score-model}. 
More recently, diffusion models have also been used for speech separation \cite{RobinSS}. However, diffusion models have not been explored yet for TSE.

TSE aims at extracting the speech of a target speaker in a multi-talker mixture\cite{zmolikova2023_spm}. Unlike other speech enhancement techniques, TSE conditions the enhancement process on a clue that identifies the target speaker in the mixture, 
such as a pre-recorded enrollment utterance of the target speaker \cite{zmolikova2017speaker,wang2018voicefilter,zmolikova2019speakerbeam,ge2020spex} or a video of the target speaker’s lip movements \cite{ephrat2018looking,afouras2018conversation,owens2018audio}. TSE is related to speech separation. However, unlike separation, which estimates all the speakers in a mixture, TSE outputs only the speech of the target speakers. This difference implies that TSE does not require estimating the number of speakers in the mixture and avoids any speaker permutation ambiguity at the output. It is thus a practical alternative to source separation for processing speech mixtures when target speaker clues are available. 

In this paper, we propose a diffusion model-based TSE. Using diffusion models for TSE requires conditioning the score model on the mixture and the target speaker clue, which can be realized using  a conditional diffusion model introduced for image generation \cite{YangSong-ICLR2021,NEURIPS2021_cfe8504b,NEURIPS2021_49ad23d1,Guidance2021neurips}. 
We explore two options to implement the score model, 1) a direct extension of the discriminative TSE, called \emph{Diff-TSE} and 2) a multi-task (MT) model that combines discriminative and Diff-TSE, called \emph{Diff-TSE-MT}.
We show experimentally that the proposed Diff-TSE and Diff-TSE-MT can extract the target speech in a mixture. However, we observe that extraction performance depends on the samples, and the system sometimes confuses the target and the interference. To resolve this issue, we exploit the generative property of the model and propose performing ensemble over multiple samples during inference. This simple approach leads to a significant extraction performance improvement. In particular, the proposed Diff-TSE-MT with ensemble inference outperforms a comparable discriminative system by a large margin.

\section{Proposed conditional diffusion model-based TSE}
\label{sec:diffTSE}
\newcommand{\bfx}{\mathbf{x}}
\newcommand{\bfv}{\mathbf{v}}
\newcommand{\bfy}{\mathbf{y}}
\newcommand{\bfz}{\mathbf{z}}
\newcommand{\bfc}{\mathbf{c}}
\newcommand{\bff}{\mathbf{f}}
\newcommand{\bfw}{\mathbf{w}}
\newcommand{\dd}{\text{d}}
\newcommand{\bfmu}{{\boldsymbol \mu}}
\newcommand{\bfI}{\mathbf{I}}
\newcommand{\bfe}{\mathbf{e}}
\newcommand{\bfa}{\mathbf{a}}

\subsection{Target speech extraction (TSE)}
\begin{figure}
    \centering\includegraphics[width=1\linewidth]{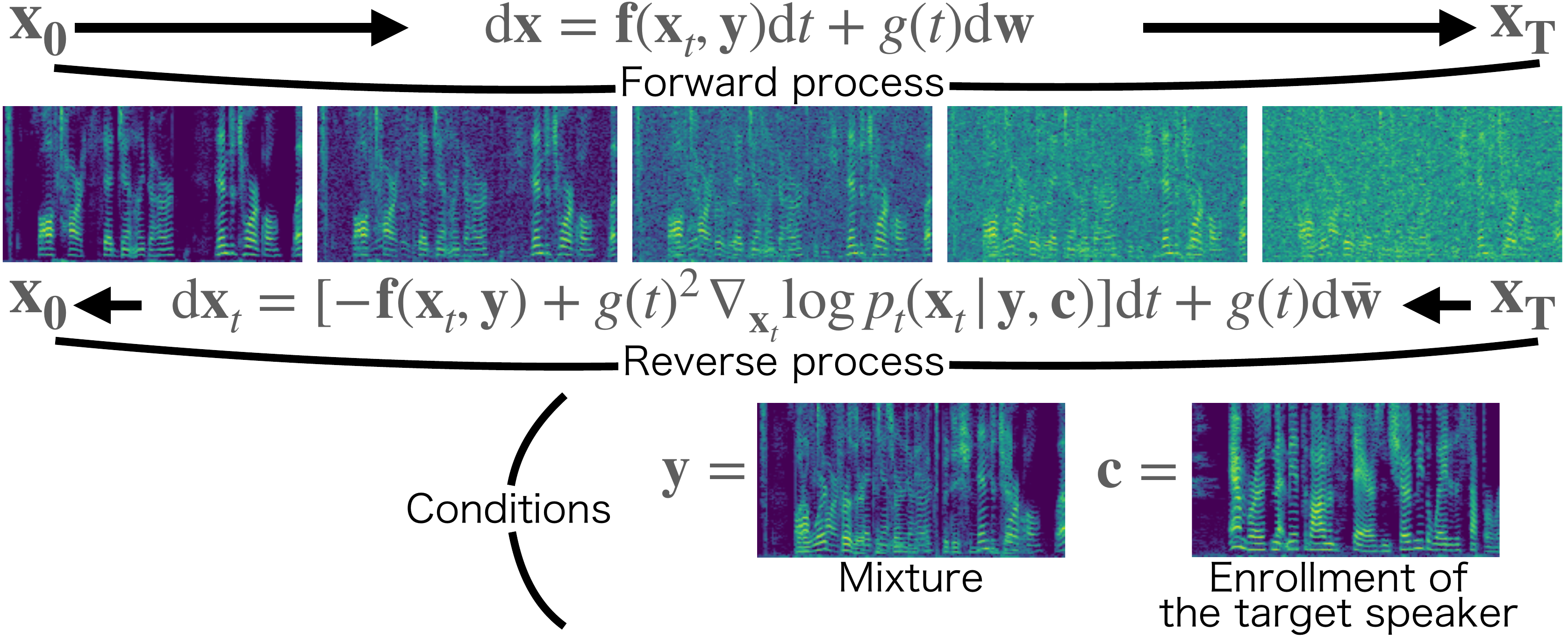}
    \vspace{-5mm}
    \caption{Conditional diffusion process conditioned on the mixture and the enrollment of the target speaker.}
    \label{fig:my_label}
    \vspace{-4mm}
\end{figure}
Let $\bfy\in\mathbb{C}^{F\times L}$ be a mixture of speech signals in the complex spectral domain, where $F$ and $L$ are the numbers of frequencies and time frames.
TSE aims to extract the speech signal uttered by a target speaker, $\bfx_0$, (target speech) from the mixture, $\bfy$, based on a clue, $\bfc$, associated with that speaker. 
Mathematically, TSE is defined as:
\begin{align}
     \hat{\bfx}_0=\mbox{TSE}(\bfy,\bfc).
\end{align}
where $\hat{\bfx}_0$ is the estimated target speech. In this paper, we use an enrollment utterance by the target speaker recorded independently of the mixture as the clue. TSE should be able to extract any speaker in the mixture associated with the clue.

Conventionally, TSE has been performed using a neural network trained in a discriminative way to minimize the errors between clean and estimated target speech signals.

\subsection{Conditional diffusion model for TSE}
This paper proposes to perform TSE in a generative way with a conditional diffusion model. The model is characterized by two stochastic differential equations (SDEs), respectively, modeling the forward and reverse  processes \cite{YangSong-ICLR2021,lu22_iccassp,Welker22interspeech} shown in Fig.~\ref{fig:my_label}. 
The forward process transforms a clean speech, $\bfx_0$, to a speech mixture corrupted with Gaussian noise.
We here assume that $\bfx_0$ follows a certain initial distribution conditioned by a speech mixture and a clue, $p(\bfx_0|\bfy,\bfc)$.
The reverse process reversely transforms a speech mixture with Gaussian noise to a clean speech that follows the initial distribution, $p(\bfx_0|\bfy,\bfc)$. 
To realize TSE, we use the forward and reverse processes, respectively, for training and inference of a TSE system. 

We adopt an SDE proposed for speech enhancement \cite{Welker22interspeech} for the forward process:
\begin{align}
    \hspace{-.5mm}\dd\bfx_t=\underbrace{\gamma(\bfy-\bfx_t)}_{\bff(\bfx_t,\bfy)}{\dd}t+\underbrace{\left[\sigma_0\left(\frac{\sigma_{1}}{\sigma_{0}}\right)^t
    \sqrt{2\log\left(\frac{\sigma_{1}}{\sigma_{0}}\right)}\right]}_{g(t)}\dd\bfw,\label{eq:forward_sde}
\end{align}
where $\bfx_t$ is the state of the process at time $t\in[0,T]$, $\bff$ and $g$ are drift and diffusion coefficient functions, and $\bfw$ is a standard Wiener process. 
$\gamma$ ($>0$) is a stiffness parameter, and $\sigma_0$ and $\sigma_1$ ($>0$) are noise scheduling parameters. This model is conditioned on the speech mixture, $\bfy$, by including it in the drift function.  This causes the forward process to move from $\bfx_0$ towards the mixture $\bfy$ by adding noise with an increased variance governed by the noise scheduling parameters. 

The solution to Eq.~\eqref{eq:forward_sde} for $\bfx_t$ follows the complex Gaussian distribution, called perturbation kernel:
\begin{align}
    p_{t}(\bfx_t|\bfx_0,\bfy) &= {\cal N}_c(\bfmu(\bfx_0,\bfy,t),\sigma(t)^2\bfI),\label{eq:pkernel}\\
    \bfmu(\bfx_0,\bfy,t) &= e^{-\gamma t}\bfx_0 + (1-e^{-\gamma t})\bfy,\label{eq:mu}\\
    \sigma(t)^2 &=\textstyle\frac{\sigma_0^2\left((\sigma_1/\sigma_0)^{2t}-e^{-2\gamma t}\right)\log(\sigma_1/\sigma_0)}{\gamma+\log(\sigma_1/\sigma_0)}.
\end{align}

To use the above model for TSE, this paper newly introduces an important assumption that differentiates the model from the conventional method \cite{Welker22interspeech}. We assume that the initial distribution of $\bfx_0$ follows $p(\bfx_0|\bfy,\bfc)$, i.e., dependent not only on the speech mixture but also on the clue. 
It makes the two SDEs aware of the distribution of the clean speech dependent on the target speaker's clue and allows us to identify which is the target speech in the mixture. It is essential for TSE.

With this assumption, we can derive the reverse SDE using Anderson's theorem
\cite{Anderson,YangSong-ICLR2021}:
\begin{align}
    \hspace{-1mm}\dd\bfx_t = [-{\bff}(\bfx_t,\bfy)+g(t)^2\nabla_{\bfx_t}\log p_t(\bfx_t|\bfy,\bfc)]{\dd}t+g(t)\dd\bar{\bfw}, \label{eq:reverse_sde2}
\end{align}
where $\bar{\bfw}$ is a standard Wiener process in reverse time and $\nabla_{\bfx_t}\log p_t(\bfx_t|\bfy,\bfc)$ 
is the gradient of $\log p_t(\bfx_t|\bfy,\bfc)$ with respect to $\bfx_t$, called a conditional score.

To perform TSE with this model, we solve the reverse SDE in Eq.~\eqref{eq:reverse_sde2} from $T$ to $0$ given $\bfx_T$, $\bfy$, and $\bfc$. 
Here, following \cite{Welker22interspeech}, we start from $\bfx_T$ sampled from the distribution shown in Eq.~\eqref{eq:pkernel} assuming $\bfmu(\bfx_0,\bfy,T)=\bfy$ instead of Eq.~\eqref{eq:mu} because $\bfx_0$ is not available for the inference. For realizing TSE, the condition $\bfc$ plays a crucial role in constraining the estimated speech 
to follow the distribution conditioned by the target speaker's clue $p(\bfx_0|\bfy,\bfc)$ and not by that of the interference speech. 

The conditional score $\nabla_{\bfx_t}\log p_t(\bfx_t|\bfy,\bfc)$ is not readily available in general, but we can use a neural network (score model)  to approximate it \cite{YangSong-ICLR2021}. 
Ideally, the score model should minimize the following loss:
\begin{align}
    \mathbb{E}_{t,(\bfx_t,\bfy,\bfc)}\left[\Vert\nabla_{\bfx_t}\log p_t(\bfx_t|\bfy,\bfc)-s_\theta(\bfx_t,\bfy,\bfc,t)\Vert_2^2\right],\label{eq:minimizer0}
\end{align}
where $s_\theta(\bfx_t,\bfy,\bfc,t)$ is the score model with a parameter set $\theta$, $\mathbb{E}_{t,(\bfx_t,\bfy,\bfc)}$ denotes the expectation over $t$ and $\bfx_t,\bfy,\bfc\sim p_t(\bfx_t,\bfy,\bfc)$. There are several techniques to train such a score model for a conditional score \cite{YangSong-ICLR2021,NEURIPS2021_cfe8504b,NEURIPS2021_49ad23d1,Guidance2021neurips}. 
Here, we use a theorem showing that minimizing the loss of Eq.~\eqref{eq:minimizer0} is equivalent to minimizing the following loss~\cite{arxiv-CDE}:
\begin{align}
    \mathbb{E}_{t,(\bfx_0,\bfy,\bfc),(\bfx_t|\bfx_0,\bfy)}\left[\Vert\nabla_{\bfx_t}\log p_t(\bfx_t|\bfx_0,\bfy)
    -s_\theta(\bfx_t,\bfy,\bfc,t)\Vert_2^2\right],\label{eq:minimizer}
\end{align}
where $\mathbb{E}_{(\bfx_t|\bfx_0,\bfy)}$ denotes expectation over $\bfx_t\sim p_t(\bfx_t|\bfx_0,\bfy)$. 
The advantage of using the loss Eq.~\eqref{eq:minimizer} is that we can calculate $\nabla_{\bfx_t}\log p_t(\bfx_t|\bfx_0,\bfy)$ in a closed form similar to conventional diffusion models \cite{YangSong-ICLR2021,Welker22interspeech}.

\subsection{Training of score model}
Using Eq.~\eqref{eq:pkernel}, the conditional score in Eq.~\eqref{eq:minimizer} becomes:
\begin{align}
    \nabla_{\bfx_t}\log p_t(\bfx_t|\bfx_0,\bfy)=-\frac{\bfx_t-\bfmu(\bfx_0,\bfy,t)}{\sigma(t)^2}.
\end{align}
Then, setting $\bfx_t=\bfmu(\bfx_0,\bfy,t)+\sigma(t)\bfz$ where $\bfz\sim{\cal N}(0,\bfI)$ based on the reparameterization trick \cite{YangSong-ICLR2021}, we can derive the score matching objective for $0\le t<T$ from the loss (\ref{eq:minimizer}):
\begin{align}
    {\cal J}^{\text{score}}(\theta)=\mathbb{E}_{t,(\bfx_0,\bfy,\bfc),\bfz}\left[\left\Vert{s_\theta(\bfx_t,\bfy,\bfc,t)+\frac{\bfz}{\sigma(t)}}\right\Vert_2^2\right].
    \label{eq:loss_score}
\end{align}
For $t=T$, because we approximate $\bfmu(\bfx_0,\bfy,\bfc,T)=\bfy$ for inference, we obtain a slightly modified objective:
\begin{align}
    {\cal J}^{\text{score}}(\theta)=\nonumber\\
    \mathbb{E}_{(\bfx_0,\bfy,\bfc),\bfz}&\left[\left\Vert{s_\theta(\bfx_T,\bfy,\bfc,T)+\frac{\bfz}{\sigma(T)}}
    {+\frac{e^{-\gamma T}(\bfx_0 -\bfy)}{\sigma(T)^2}}\right\Vert_2^2\right],\label{eq:robin}
\end{align}
which corresponds to adding loss minimizing the distance between $\bfmu(\bfx_0,\bfy,\bfc,T)$ obtained with Eq.~\eqref{eq:mu} and $\bfy$. A similar objective was introduced for diffusion model-based source separation \cite{RobinSS} except that we do not need to consider permutation errors in the loss for TSE.

In practice, we calculate the expectation of the above objectives by their average over data randomly sampled from a training dataset. 
At each sampling, we first sample a target speaker and then sample the data fixing the target speaker, implicitly assuming that $\bfc$ uniquely determines the target speaker.
We set the probability to sample $t=T$ and $0\le t<T$ at $\delta_T$ and $1-\delta_T$ to put a relatively large weight to Eq.~(\ref{eq:robin}).

\subsection{Configurations for the score model}
\begin{figure}
    \centering\includegraphics[width=1.0\linewidth]{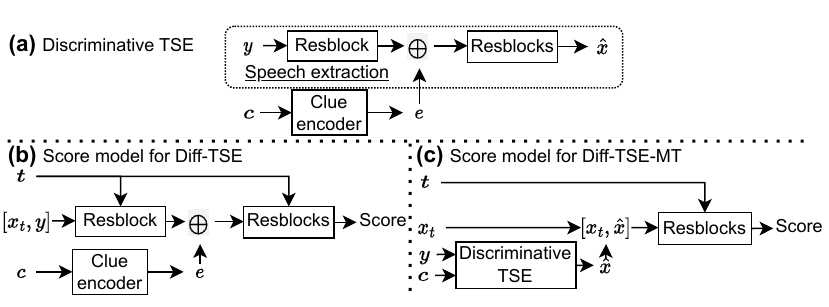}
    \vspace{-5mm}
    \caption{Schematic diagrams of (a) a discriminative TSE system, the score model of the proposed (b) Diff-TSE and (c) Diff-TSE-MT. Resblock indicates the residual block defined in NCSN++. We use an enrollment utterance as the clue, $\bfc$.}
    \label{fig:net_config}
    \vspace{-5mm}
\end{figure}
We explore different configurations for the score model, $s_\theta(\bfx_t,\bfy,\bfc,t)$. These configurations are related to the network architecture of discriminative TSE systems and differ mainly by how we implement the conditioning on the clue.

First, we briefly review a typical discriminative TSE system, consisting of a clue encoder network and a speech extraction network, as shown in Fig.~\ref{fig:net_config}-(a). The clue encoder network computes a target speaker embedding vector, $\bfe $, from an enrollment utterance, $\bfc$, using a simple network with an average pooling layer as the output layer. The extraction network accepts the mixture and controls the extraction on the target speaker embedding, using a fusion layer internally. It consists of a stack of residual blocks. We use here the multiplication fusion layer \cite{Delcroix_ICASSP2019}. The output of the extraction network is the estimated target speech, $\hat{\bfx}_0$. We train a discriminative TSE system by optimizing the SNR between clean and estimated target speech, i.e., ${\cal J}^{\text{SNR}}(\theta^{\text{TSE}}) = - \text{SNR}(\bfx_0,\hat{\bfx}_0)$.

We can use a similar configuration to the TSE system for the score model. We call this approach \emph{Diff-TSE}.  It is shown in Fig. \ref{fig:net_config}-(b). We use a similar clue encoder network to process the enrollment utterance as discriminative TSE. The differences with the discriminative TSE system are that the input of the extraction network consists of the concatenation of $\bfx_t$ and the mixture $\bfy$, and the output is the score. Besides, it is trained using the loss of Eqs.~\eqref{eq:loss_score} and \eqref{eq:robin}. 

As an alternative, we also explore an MT model combining discriminative TSE and Diff-TSE shown in Fig. \ref{fig:net_config}-(c). We call it \emph{Diff-TSE-MT}.
The input of the score model is the same as that of Diff-TSE (i.e., $\bfx_t$, $\bfy$, $\bfc$), but we put $\bfy$ and $\bfc$ into the discriminative TSE model to estimate the target speech $\hat{\bfx}_0$ internally.  We then concatenate $\hat{\bfx}_0$ with $\bfx_t$ and put it into a subsequent network to estimate the score.
The score model, including the discriminative TSE model, is trained using an MT objective:
\begin{align}
{\cal J}^{\text{MT}}(\theta^{\text{TSE}}, \theta^{\text{score}}) = \alpha {\cal J}^{\text{SNR}}(\theta^{\text{TSE}}) + \beta{\cal J}^{\text{score}}(\theta^{\text{TSE}}, \theta^{\text{score}}),
\label{eq:multi_task_objective}
\end{align}
where $\alpha$ and $\beta$ are multi-task weights. ${\cal J}^{\text{SNR}}(\theta^{\text{TSE}})$ is computed on the output of the discriminative TSE model. $\theta^{\text{TSE}}$ and $ \theta^{\text{score}}$ are the parameters of the TSE module and the extraction network, respectively. Note that we train all network parameters jointly from scratch, as for the other configurations.

\subsection{Inference with ensemble}
The inference process consists of running the reverse diffusion process of Eq.~\eqref{eq:reverse_sde2}, approximating the conditional score $\nabla_{\bfx_t}\log p_t(\bfx_t|\bfy,\bfc)$ by the output of the score model $s_\theta(\bfx_t,\bfy,\bfc,t)$. It is solved
by the reverse sampling based on the Predictor-Corrector sampler \cite{YangSong-ICLR2021}, 
starting from $\bfx_T \sim {\cal N}_c(\bfy,\sigma(T)^2\bfI)$, to $t\approx 0$. 
Note that, unlike discriminative TSE, Diff-TSE provides access to the distribution of the target speech given the enrollment and the mixture. We propose to exploit this property of Diff-TSE to improve extraction performance. 

In our preliminary experiments, we observed that some extracted samples had segments where the interference speaker was extracted instead of the target. We hypothesize that these samples may be outliers from the posterior distribution. Therefore, to mitigate their impact, we propose performing ensemble over several extracted samples obtained by repeating the inference process several times with different random seeds. We obtain the extracted speech after ensemble as,
$
\hat{\bfx}_0^{\text{Ens}} = \sum_j^J \hat{\bfx}_0^j,$
where $j$ is the sample index.

\section{Related work}
\label{sec:related}
Recently, researchers have proposed techniques to condition diffusion model-based speech enhancement on a pre-processed speech obtained by another speech enhancement method. For example, Universal Speech Enhancement (USE)~\cite{Dolby-score-model} uses pre-processed speech as a condition of the score model to decide what signal to generate from Gaussian noise. 
Stochastic Regeneration Model (StoRM)~\cite{StoRM} regenerates an improved enhanced speech from a pre-processed one using a reverse SDE similar to Eq.~(\ref{eq:reverse_sde2}). 

In contrast, our proposed method, Diff-TSE-MT, provides a different way to use the pre-processed speech in the model. While our method internally estimates and uses the pre-processed speech, it still uses the observed speech to condition the diffusion model. An advantage of our method is that the model can use not only the pre-processed speech but also the observed speech to generate the target speech, which can cover information that may be missed in the pre-processed speech. Future work should include an experimental comparison of our method with USE and StoRM-like conditioning schemes.

\section{Experiments}
\label{sec:expe}
We perform experiments using the openly available LibriMix-2spk dataset~\cite{cosentino2020librimix}. We use the 100 version of the data and follow the openly available recipe\footnote{\label{fn:katka}\url{https://github.com/butspeechfit/speakerbeam}}, which defines the enrollment utterances used for each mixture in the test set. 

\subsection{Settings}
We based our implementation of Diff-TSE on the publicly available code for diffusion-based speech enhancement,\footnote{\url{https://github.com/sp-uhh/sgmse}} which we adapted for the TSE task.
We use the Noise Conditional Score Network (NCSN++) architecture \cite{YangSong-ICLR2021,Julius22arxiv} for TSE, Diff-TSE, and Diff-TSE-MT models. The main difference is that the TSE and Diff-TSE insert the fusion layer after the first residual block of the NSCN++ network. We use three BLSTM layers with 1024 hidden units followed by an average pooling layer for the clue encoder network.
We use short-time Fourier transform (STFT) coefficients with transformed amplitude as in \cite{Julius22arxiv}.

We train all networks using the Adam optimizer with a learning rate $lr = \num{1e-4}$ and exponential averaging of the network weights \cite{Julius22arxiv}. For Diff-TSE and Diff-TSE-MT training, we use 
$\delta_{T} = 0.1$ for sampling $t=T$.
We use $\alpha=\beta=1.0$ for multi-task loss in Eq.~(\ref{eq:multi_task_objective}).
For the parameters of SDE in Eq.~\eqref{eq:forward_sde}, we set $\gamma=2$, $\sigma_0 = 0.05$, and $\sigma_1 = 0.5$.

For inference, we use a number of prediction steps $N=30$ and a step size $r = 0.5$. We use ten samples, i.e., $J=10$, when performing ensemble inference. We measure the performance in terms of the perceptual evaluation of speech quality (PESQ)~\cite{PESQ}, the extended short-time objective intelligibility (ESTOI)~\cite{ANDERSEN20181}, and the scale-invariant signal-to-distortion ratio (SI-SDR)~\cite{le2019sdr}.

\subsection{Results}
\begin{table}[tb]
  \caption{TSE performance for different models. D/G indicates if the model is discriminative or generative.}
  \vspace{-3mm}
  \label{tab:results}
  \centering
  \begin{tabular}{llcccc }
    \toprule
    &Model &  D/G & PESQ & ESTOI & SI-SDR\\
    \midrule
    0 & Mixture & - & 1.60 & 0.54 & 0.03 \\
    \midrule
    1& TSE  & D & 2.58 &  0.75 & 10.01 \\
    \midrule
   2& Diff-TSE & G &  2.56& 0.74 & 7.85\\
   3& +Ensemble & G &  2.90& 0.76 & 9.49\\
    \midrule
    4 & Diff-TSE-MT & D & 2.60& 0.76 & 10.71  \\
    5 & Diff-TSE-MT & G & 2.79 & 0.77 &  9.40   \\
    6 & +Ensemble  & G & 3.08  & 0.80 & 11.28   \\
    \bottomrule
  \end{tabular}
  \vspace{-3mm}
\end{table}

Table~\ref{tab:results} compares the performance of the discriminative and diffusion-based TSE systems. 
The results demonstrate that performing TSE with a diffusion model is possible. Diff-TSE without ensemble inference (system 2) achieves comparable PESQ and ESTOI but much lower SI-SDR than the discriminative TSE system (system 1).
However, the performance of Diff-TSE improves with the ensemble inference (system 3).

The Diff-TSE-MT systems (systems 5 and 6) achieves superior performance than the Diff-TSE model (system 2 and 3). 
Interestingly, the discriminative TSE module within Diff-TSE-MT (system 4) outperforms the generative Diff-TSE-MT (system 5
) in terms of SI-SDR but performs slightly worse in terms of PESQ and ESTOI. Diff-TSE-MT with ensemble inference (system 6) outperforms all other systems for all metrics.

These numbers are not necessarily state-of-the-art. For example, time-domain SpeakerBeam\footref{fn:katka} achieves PESQ, ESTOI, and SI-SDR values of 2.76, 0.81, and 12.84~dB, respectively, on the same task. In this preliminary study, we employed the NSCN++ architecture, which was successful for diffusion model-based noise reduction and dereverberation but may not be optimal for speech separation\cite{RobinSS} or TSE. 
The current implementation of Diff-TSE-MT+Ensemble performs thus worse in terms of SI-SDR, but better in terms of PESQ. These results demonstrate the potential of using diffusion models for TSE and the importance of the proposed ensemble inference.
Our future work will investigate better network architectures for Diff-TSE.

\subsection{Analysis of the ensemble inference method}

\begin{figure}[t]
  \centering
\includegraphics[width=1.0\linewidth]{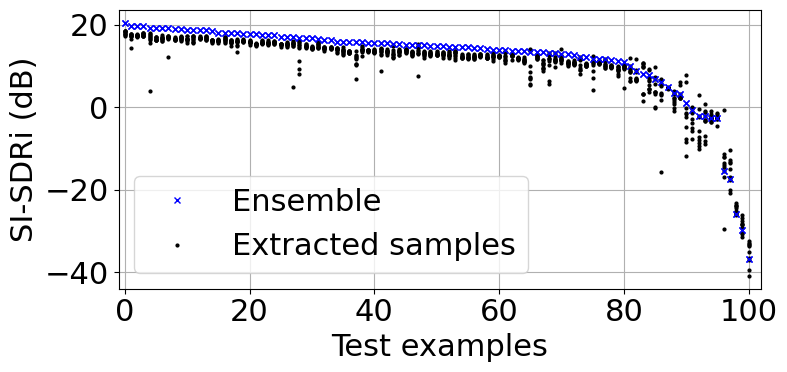}
\vspace{-4mm}
  \caption{SI-SDRi for 100 randomly selected test examples, showing the results of 10 extracted samples with Diff-TSE-MT (black dots), and the ensemble inference (blue crosses).}
  \label{fig:ensemble_inference}
  \vspace{-3mm}
\end{figure}

\begin{figure}[t]
\centering
\begin{minipage}{.45\linewidth}
  \centering
  \includegraphics[width=1\linewidth]{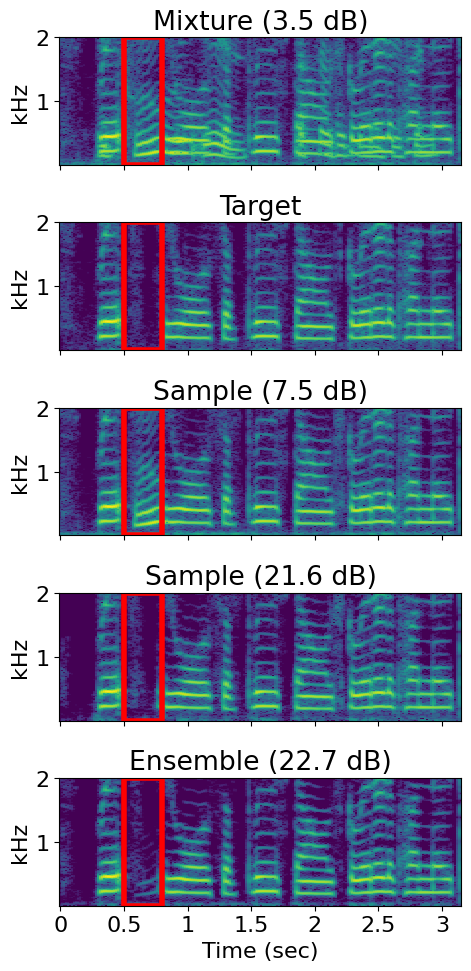}
\end{minipage}%
\begin{minipage}{.45\linewidth}
  \centering
  \includegraphics[width=1\linewidth]{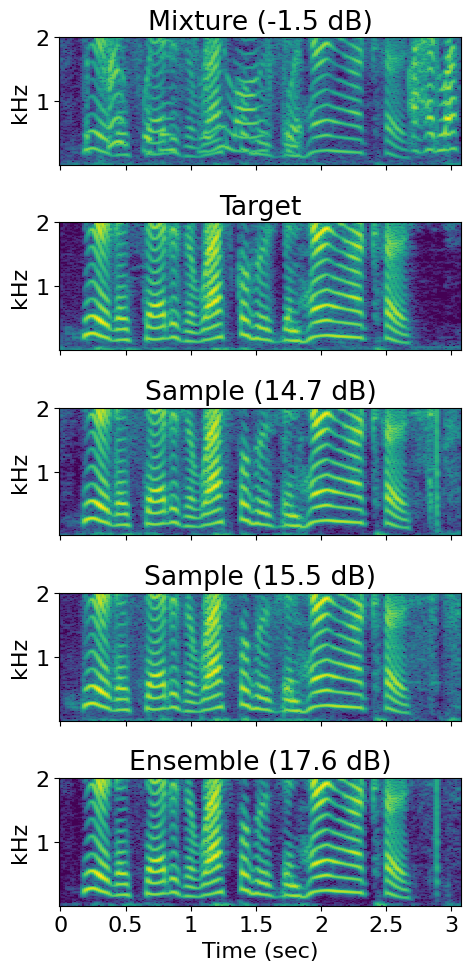}
\end{minipage}

\vspace{-4mm}
\caption{Spectrogram for two representative test examples showing the worst and best of the 10 samples generated with Diff-TSE-MT, and the results of ensemble inference. The numbers in parentheses are the SI-SDR.} 
\label{fig:spectrograms}
\vspace{-5mm}
\end{figure}

Figure \ref{fig:ensemble_inference} shows the SI-SDR improvement (SI-SDRi) for 100 randomly selected test examples processed with Diff-TSE-MT. We generated 10 extracted samples for each test example and computed the ensemble. 
Interestingly, the ensemble inference often performs better than all of the individual extracted samples, even when all samples had high SI-SDR (above 10 dB).

Figure~\ref{fig:spectrograms} shows the spectrograms of two representative text examples, showing samples with the worst and best SI-SDR and the ensemble result. In the left figure, the worst sample includes extraction errors shown in the red box. However, other samples perform better. With ensemble inference, we can mitigate the impact of the extraction error. The right figure shows an example where all samples had high SI-SDR. Still, even in this case, ensemble inference reveals to be effective in further improving extraction performance by mitigating relatively minor artifacts. 

These results indicate that ensemble inference does not only mitigate the influence of obvious extraction errors but also improves performance in general. However, unsurprisingly, ensemble inference does not help when all samples have a low SI-SDR (below 0 dB). These samples are often complete extraction failures, where TSE extracts mostly the interference speaker.

\section{Conclusions}
\label{sec:conclusion}
In this paper, we proposed a diffusion model-based TSE system implemented using a conditional diffusion model. We presented two approaches for implementing the score model and introduced an ensemble inference scheme to mitigate extraction errors. We showed that the proposed diffusion model-based TSE outperforms a comparable discriminative TSE model in terms of PSEQ, ETOI, and SI-SDR.

In our future work, we will investigate approaches to reduce extraction failures of Diff-TSE by exploring other network configurations \cite{ge2020spex} and generating more discriminative speaker embeddings \cite{delcroix2020improving,mun2020sound}.

\bibliographystyle{IEEEtran}
\bibliography{mybib}

\end{document}